\documentclass[%
 reprint,
nofootinbib,
 amsmath,amssymb,
 aps,
 prl,
]{revtex4-2}

\usepackage[normalem]{ulem}
\usepackage{multirow}
\usepackage{graphicx}
\usepackage{dcolumn}
\usepackage{bm}
\usepackage[dvipsnames]{xcolor}
\usepackage{hyperref}
\usepackage{enumerate}
\usepackage{amsmath}
\usepackage{CJKutf8}

\newcommand{\UM}{Department of Physics, University of Michigan, 450 Church St, Ann Arbor, MI 48109, USA}
\newcommand{\LCTP}{Leinweber Center for Theoretical Physics, 450 Church St, Ann Arbor, MI 48109, USA}

\begin{document}

\preprint{APS/123-QED}

\title{Ten-dimensional neural network emulator for the nonlinear matter power spectrum}

\author{\begin{CJK}{UTF8}{gbsn}Yanhui Yang (\CJKfamily{gbsn}杨焱辉)$^1$\end{CJK}}

 \email{yyang440@ucr.edu}
\author{Simeon Bird$^1$}%
 \email{sbird@ucr.edu}
\author{\begin{CJK}{UTF8}{bsmi}Ming-Feng Ho (何銘峰)$^{1,2,3}$\end{CJK}}

\author{Mahdi Qezlou$^4$}

\affiliation{%
$^1$Department of Physics \& Astronomy, University of California, Riverside, 900 University Ave., Riverside, CA 92521, USA
}%
\affiliation{$^2$\UM}
\affiliation{$^3$\LCTP}
\affiliation{$^4$The University of Texas at Austin, 2515 Speedway Boulevard, Stop C1400, Austin, TX 78712, USA
}




\date{\today; published in Phys. Rev. Lett. on February 9, 2026}

\begin{abstract}

We present {\scriptsize GokuNEmu}, a ten-dimensional neural network emulator for the nonlinear matter power spectrum, designed to support next-generation cosmological analyses. Built on the {\scriptsize Goku} $N$-body simulation suite and the {\scriptsize T2N-MusE} emulation framework, {\scriptsize GokuNEmu} predicts the matter power spectrum with $\sim 0.5 \%$ average accuracy for redshifts $0 \leq z \leq 3$ and scales $0.006 \leq k/(h\,\mathrm{Mpc}^{-1}) \leq 10$. The emulator models a 10D parameter space that extends beyond the $\Lambda$-cold dark matter ($\Lambda$CDM) model to include dynamical dark energy (characterized by $w_0$ and $w_a$), massive neutrinos ($\sum m_\nu$), the effective number of neutrinos ($N_\text{eff}$), and running of the primordial spectral index ($\alpha_\text{s}$). Its broad parameter coverage, particularly for the extensions, makes it the only matter power spectrum emulator encompassing the range of dynamical dark energy models preferred by recent DESI constraints. In addition, it requires only $\sim $2 milliseconds to predict a single cosmology on a laptop, orders of magnitude faster than existing emulators. These features make {\scriptsize GokuNEmu} a uniquely powerful tool for interpreting observational data from upcoming surveys such as LSST, \textit{Euclid}, the Roman Space Telescope, and CSST.

\end{abstract}

\maketitle


\section{\label{sec:intro}Introduction}

Cosmological surveys \cite[e.g.~][]{DESICollaboration2016, LSC2009, Laureijs2011,Akeson2019,Gong2019, Takada2014} constrain cosmological models motivated by unresolved fundamental physics questions, such as the accelerated expansion of the Universe~\cite{Caldwell2009}, the nature of dark matter (DM)~\cite{Feng2010}, the sum of neutrino masses~\cite{Wong2011}, and the parameter tensions found in $\Lambda$CDM \cite[e.g.~][]{Riess2021,Riess2022,Asgari2021,Abbott2022}.
Interpreting these measurements requires an accurate understanding of how the underlying cosmological parameters affect the growth of large-scale structure  in the Universe. Pre-trained simulation-based cosmological emulators which predict summary statistics are used for this purpose, obviating the need for intensive numerical computation at every likelihood evaluation~\cite[e.g.~][]{Heitmann2009,Heitmann2010,Heitmann2013, DeRose2019,McClintock2019,Zhai2019, Smith2019, Nishimichi2019, Valcin2019,Arico2021, Villaescusa2020, Heitmann2016,Lawrence2017,Bocquet2020,Moran2023,Kwan2023,Casares2024, Knabenhans2019,Knabenhans2021, Chen2025,Chen2025b, Yang2025,Cabayol-Garcia2023,Diao2025}. However, none of the existing emulators (except {\scriptsize GokuEmu}~\cite{Yang2025}) cover a wide enough prior parameter range to match recent dark energy (DE) constraints~\cite{Abdul-Karim2025,Adame2025,Abbott2025}, nor are they scalable to higher dimensional parameter spaces due to the high cost of running high-fidelity training simulations. Here we present a new emulator, \texttt{GokuNEmu}, a highly accurate and computationally efficient neural network model which achieves uniquely broad coverage of parameter and redshift space by using the {\scriptsize Goku} simulations~\cite{Yang2025}.

The {\scriptsize Goku} simulations were the first to expand the parameter space to 10 dimensions, including the $5$-parameter base $\Lambda$CDM model and several extensions which directly impact the matter power spectrum. These are: dynamical dark energy (characterized by $w_0$ and $w_a$), massive neutrinos ($\sum m_\nu$), the effective number of neutrinos ($N_\text{eff}$), and running of the spectral index ($\alpha_\text{s}$).
The {\scriptsize Goku} suite covers a uniquely wide prior range of dynamical DE parameters, making it the \textit{only} simulation suite to safely cover recent constraints from the Dark Energy Spectroscopic Instrument (DESI)~\cite{Abdul-Karim2025,Adame2025} and the Dark Energy Survey (DES)~\cite{Abbott2025}. {\scriptsize Goku} uses a multifidelity (MF) strategy \cite{Ho2022,Ho2023}, where a large number of lower cost simulations explore parameter space and are corrected by a small number of high cost, more accurate, simulations, reducing the total computational cost by 94\%~\cite{Yang2025}. The simulations were used to build a Gaussian process (GP) emulator for the nonlinear matter power spectrum, called {\scriptsize GokuEmu}~\cite{Yang2025}.

We present {\scriptsize GokuNEmu}, a neural network (NN) emulator trained on {\scriptsize Goku} for the nonlinear matter power spectrum using {\scriptsize T2N-MusE}, an MF emulation framework we developed for training highly optimized fully-connected neural networks (FCNNs)~\cite{Yang2025b}. While GP regression has been widely used in cosmological emulation, NNs offer greater efficiency in both training and inference, particularly in high-dimensional parameter spaces and with large datasets. Given that the {\scriptsize Goku} simulations consist of 1128 pairs of low-fidelity (LF) simulations and 36 high-fidelity (HF) simulations,\footnote{Existing emulators were typically trained on $\sim 100$ samples.} NNs are a more suitable choice in this case.

{\scriptsize T2N-MusE} enables efficient training of {\scriptsize GokuNEmu} on 34 redshift bins between 0 and 3, ensuring more accurate interpolation of the matter power spectrum in redshift space than its predecessor (which included 6 redshifts). It also allows us to cover a broader range of spatial scales, from $0.006h/\mathrm{Mpc}$ to $10h/\mathrm{Mpc}$, by combining two MF models trained on low- and high-$k$ ranges respectively.\footnote{The emulation technique used in {\scriptsize GokuEmu} did not allow covering the full spatial range allowed by the simulations.} Also, {\scriptsize GokuNEmu} achieves a significantly better generalization accuracy and is orders of magnitude more efficient in evaluation time than other emulators. {\scriptsize GokuNEmu} will be especially useful for surveys that measure structures down to nonlinear scales, such as LSST~\cite{LSC2009}, \textit{Euclid}~\cite{Laureijs2011}, the Roman Space Telescope~\cite{Akeson2019}, and CSST~\cite{Gong2019}.
 
\section{\label{sec:methods}Methods}

This section describes the data and methods used to build the emulator {\scriptsize GokuNEmu}. We quantify the generalization performance of the emulator using leave-one-out cross-validation (LOOCV). HF samples are iteratively excluded from the training set and used as test points. The leave-one-out (LOO) error is defined as the relative mean absolute error (rMAE) of the emulator prediction with respect to the matter power spectrum measured from the corresponding simulation, denoted as $\Phi_\text{rMAE}$.

\subsection{\label{sec:sims}{\scriptsize Goku} Simulations}
\begin{figure}
    \includegraphics[width=\columnwidth]{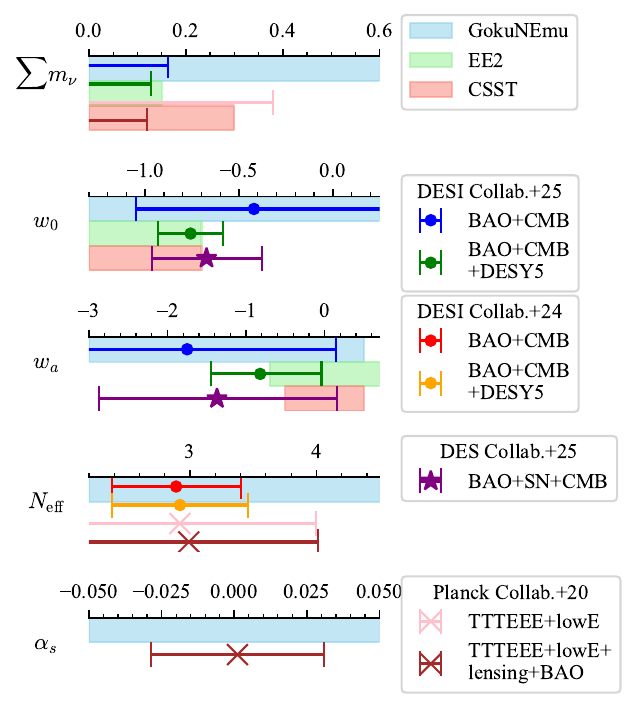}
    \caption{\label{fig:prior_compare}Prior ranges of the $\Lambda$CDM extensions of different emulators and selected cosmological constraints from the literature. The sky blue, light green, and salmon bars represent the prior ranges of {\scriptsize GokuNEmu} (this work), {\scriptsize EuclidEmulator2}, and the {\scriptsize CSST Emulator}, respectively. The constraints are shown with error bars denoting $\pm 3\sigma$\footnote{Here, $\sigma$ is defined from the 68\% credible interval.} uncertainties (though $95\%$ upper limits are shown for $\sum m_\nu$). The constraints are selected from DESI~\cite{Abdul-Karim2025,Adame2025}, DES~\cite{Abbott2025}, and \textit{Planck}~\cite{Aghanim2020}.}
\end{figure}

We briefly describe the {\scriptsize Goku} simulations here (for more details, see Ref.~\cite{Yang2025}). The {\scriptsize Goku} simulations are a suite of $N$-body simulations run with the {\scriptsize MP-Gadget} code~\cite{Feng2018}. {\scriptsize Goku}'s cosmologies are sampled using a Sliced Latin Hypercube Design~\cite{Ba2015}. An adaptive sampling strategy is used to sample the parameter space, which consists of two Latin hypercubes: a small hypercube with narrow ranges of cosmological parameters, {\scriptsize Goku-N}, and a large hypercube, {\scriptsize Goku-W}. Assuming a flat Universe, {\scriptsize Goku} covers 10 cosmological parameters: the base $\Lambda$CDM parameters including the matter density $\Omega_\text{m}$, the baryon density $\Omega_\text{b}$, the dimensionless Hubble parameter $h$, the primordial spectral index $n_\text{s}$, and the amplitude of the primordial power spectrum $A_\text{s}$, the parameters for time-dependent DE: $w_0$ and $w_a$~\cite{Chevallier2001,Linder2003}, the sum of neutrino masses $\sum m_\nu$~\cite{Ali2013,Bird2018}, the effective number of ultrarelativistic neutrinos $N_\text{eff}$~\cite{Lesgourgues2006,Shvartsman1969,Steigman1977}, and the running of the spectral index $\alpha_\text{s}$. The ranges of the parameters are listed in Table~\ref{tab:ranges}. Fig.~\ref{fig:prior_compare} shows the prior ranges for $\Lambda$CDM extensions in {\scriptsize Goku} (thus the emulator {\scriptsize GokuNEmu}) compared to those of other emulators and recent constraints from DESI~\cite{Abdul-Karim2025,Adame2025}, DES~\cite{Abbott2025}, and \textit{Planck}~\cite{Aghanim2020}. We observe that many of the constraints on $w_0$ and $w_a$ are out of the capabilities of the other emulators, while {\scriptsize GokuNEmu} covers them safely. Also, {\scriptsize EE2} struggles to cover some of the upper limits of $\sum m_\nu$, while {\scriptsize Goku}'s prior range is more than sufficient and  notably includes the direct neutrino-mass measurement from the {\scriptsize KATRIN} experiment~\cite{Aker2025} (which is independent of cosmology), $\sum m_\nu < 0.45\,\text{eV}$ (90\% confidence level). As for $N_\text{eff}$ and $\alpha_\text{s}$, {\scriptsize GokuNEmu} is also able to test the results from \textit{Planck}~\cite{Aghanim2020}.

\begin{table}
    \caption{\label{tab:ranges}%
    Parameter boxes for \texttt{Goku-W} and \texttt{Goku-N} defined through their lower  and upper bounds.}
    \begin{ruledtabular}
    \begin{tabular}{lcccc}
    \textrm{Parameter}&
    \textrm{min(\texttt{W})}& \textrm{min(\texttt{N})}& \textrm{max(\texttt{N})}&
    \textrm{max(\texttt{W})}\\
    \colrule
    $\Omega_\text{m}$ & $0.22$ &$0.26$ &$0.35$ & $0.40$\\
    $\Omega_\text{b}$& $0.040$ &$0.045$&$0.051$ & $0.055$\\
    $h$ & $0.60$ &$0.64$& $0.74$ &$0.76$\\
    $A_\text{s}$ & $1.0\times 10^{-9}$
    &$1.7\times 10^{-9}$& $2.5\times 10^{-9}$& $3.0\times 10^{-9}$\\
    $n_\text{s}$ & $0.80$ & $0.95$ & $1.00$ &$1.10$\\
    $w_0$ & $-1.30$ & $-1.30$ & $-0.70$ &$0.25$\\
    $w_a$ & $-3.0$& $-1.0$ & $0.5$ & $0.5$\\
    $\sum m_\nu$ & $0.00$ & $0.06\,\text{eV}$ & $0.15\,\text{eV}$ &$0.60\,\text{eV}$\\
    $N_\text{eff}$ & $2.2$& $2.3$ & $3.7$ & $4.5$\\
    $\alpha_\text{s}$ & $-0.05$ &$-0.03$ &$0.03$ &$0.05$\\
    \end{tabular}
    \end{ruledtabular}
\end{table}

The {\scriptsize Goku-W} ({\scriptsize -N}) set consists of 564 pairs of LF simulations and 21 (15) HF simulations. Each HF simulation evolves $3000^3$ particles within a comoving volume of $(1000\,\mathrm{Mpc}/h)^3$. The LF simulations are divided into two types: L1 and L2. Both L1 and L2 simulations include $750^3$ particles, significantly fewer than the HF runs. While L1 simulations share the same box size as the HF simulations, L2 simulations are run in smaller volumes with side lengths of $250\,\mathrm{Mpc}/h$. In this setting, the L1 simulations are used to cover the low-$k$ (large-scale) range, while the L2 simulations are more accurate in the high-$k$ (small-scale) range (see Fig.~7 of Ref.~\cite{Yang2025}). The LF samples in emulation are responsible for interpolation (exploring the parameter space), while the HF samples are used to correct the bias of the LF samples.

\subsection{\label{sec:sim_data}Data and Combination of Emulators}

For each of the simulations, in addition to the 6 fixed redshift bins at $z=0$, $0.2$, $0.5$, $1$, $2$, and $3$, the matter power spectrum was also computed and saved at $\gtrsim 100$ redshifts between $z=0$ and $z=3$. We interpolate the matter power spectrum at 28 extra $z$ bins within $0<z<3$, evenly spaced in $\ln a$ (where $a$ is the scale factor), and combine them with the original 6 redshifts to form a total of 34 redshift bins. We have confirmed that these 34 $z$ bins are sufficient to recover the original output of the simulations, with negligible interpolation error $\lesssim 10^{-3}$.

\begin{figure}
    \includegraphics[width=\linewidth]{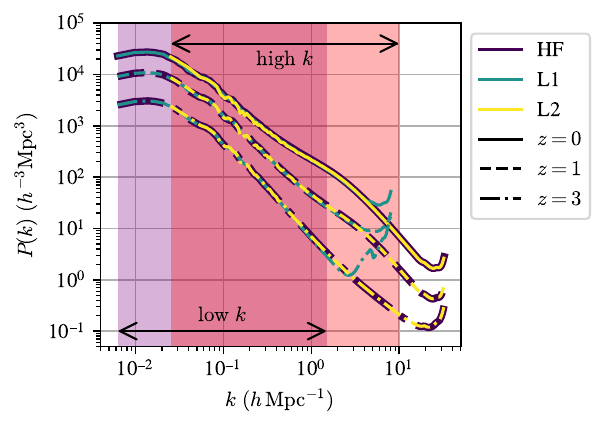}
    \caption{\label{fig:low_high_k}The matter power spectra measured from the {\scriptsize Goku-N-0195} (see Table~V of Ref.~\cite{Yang2025} for the values of the cosmological parameters) simulations at $z=0, 1$, and $3$ (solid, dashed, and dot-dashed lines, respectively). The HF, L1 and L2 power spectra are shown in different colors. The purple and red shaded regions indicate the low- and high-$k$ ranges, respectively.}
\end{figure}

We build two separate emulators for low- and high-$k$ scale ranges.
These emulators cover $0.006 < k_\text{low}/(h\,\text{Mpc}^{-1}) < 1.5$ and $0.025 < k_\text{high}/(h\,\text{Mpc}^{-1}) < 10$, respectively. As shown in Fig.~\ref{fig:low_high_k}, the L1 simulation matches the HF simulation well at large scales but deviates from it at $k\gtrsim 1.5h/\text{Mpc}$ significantly due to insufficient resolution. The L2 simulation has a box size limiting it to $k > 0.025h/\text{Mpc}$, but is a good approximation to the HF simulation on small scales. We trim the range of data used to train each emulator, as including poorly converged scales can degrade the performance of the emulator across the entire spatial range. For the low-$k$ emulator (hereafter, {\scriptsize Emu1}), the L1 and HF power spectra are truncated at $k_\text{low,max}=1.5h/\text{Mpc}$ before being used for training. For the high-$k$ emulator (hereafter, {\scriptsize Emu2}), the L2 and HF power spectra are truncated at $k_\text{high,min}=0.025h/\text{Mpc}$ and $k_\text{high,max}=10h/\text{Mpc}$ before serving as training data.

Our low- and high-$k$ emulators predict overlapping scales.
We blend the two emulators through a smooth ramp function to avoid discontinuities. Specifically, between $k_\text{min} = 0.025h/\text{Mpc}$ and $k_\text{max} = 1.5h/\text{Mpc}$, the prediction of the final emulator is given by
\begin{equation}
    P(k, z) = [1 - w(k)] \cdot P_\text{Emu1}(k, z) +  w(k) \cdot P_\text{Emu2}(k, z),
\end{equation}
where $w(k)$ is the weight function. We define $w(k)$ as a sigmoid function, i.e.,
\begin{equation}
    w(k) = \frac{1}{1 + \exp\left[-s\cdot \frac{k - k_\text{mid}}{k_\text{max} - k_\text{min}}\right]}
\end{equation}
where $k_\text{mid} = (k_\text{min} + k_\text{max}) / 2$, and $s = 4$ is a scaling factor that controls the steepness of the transition, chosen to minimize validation error.\footnote{The choice, $s=4$, also ensures that the combined prediction $P(k, z)$ remains smooth at the transition boundaries.}

Following Ref.~\cite{Yang2025}, we apply a universal pairing-and-fixing~\cite{Angulo2016} correction function to the final emulator output to mitigate cosmic variance.

\subsection{\label{sec:emu_muse}Emulation with {\scriptsize T2N-MusE}}

The {\scriptsize T2N-MusE} framework builds highly optimized MF emulators based on FCNNs and is fully described in the companion paper \cite{Yang2025b}. 
The NNs have a 2-step architecture, in which step 1 is an LF model and step 2 corrects the LF output via learning the mapping between the input (cosmological parameters) and the ratio of the HF output to the LF output. A 2-stage Bayesian optimization process is used to optimize the hyperparameters of the NNs: the number of layers, the number of neurons in each layer, and the strength of L2 regularization (which prevents overfitting). 
The first stage is a coarse search over a wide range of hyperparameters, while the second stage is a fine search within a smaller space. To train and validate the LF model, a 2-phase strategy is used, which explores the random seed space with considerably higher efficiency. In brief, we use {\scriptsize T2N-MusE} to build {\scriptsize Emu1} and {\scriptsize Emu2} separately. For more details of {\scriptsize T2N-MusE}, we refer the reader to Ref.~\cite{Yang2025b}.

\section{\label{sec:results}Results}


\begin{figure}
    \includegraphics[width=\linewidth]{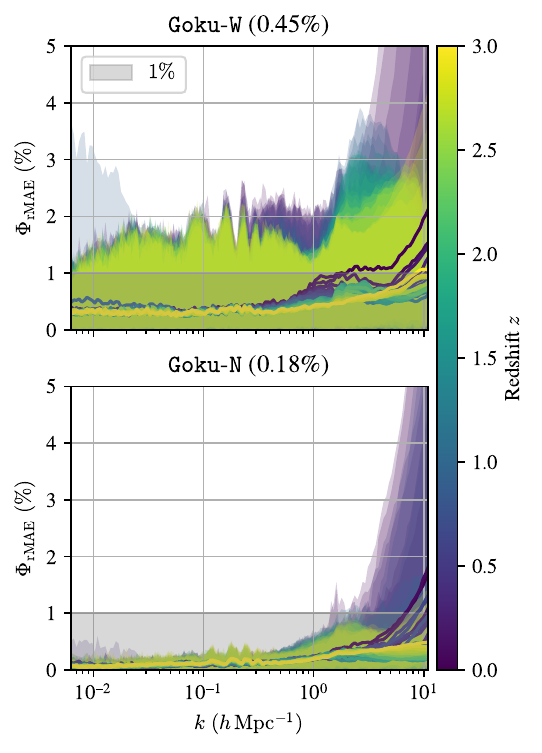}
    \caption{\label{fig:loo_WN}LOO errors of {\scriptsize GokuNEmu} in the \texttt{Goku-W} (top) and \texttt{Goku-N} (bottom) parameter spaces. Redshifts are color coded. The solid lines are the error averaged over cosmologies, and the corresponding shaded regions indicate the range of individual cosmologies. The gray-shaded areas mark the region where the error is less than 1\%. For each space, the mean error averaged over all the test points, 34 redshifts and 137 $k$ bins is shown in the title of the corresponding panel.}
\end{figure}

Fig.~\ref{fig:loo_WN} shows the LOOCV results of {\scriptsize GokuNEmu} in the {\scriptsize Goku-W} (top) and {\scriptsize Goku-N} (bottom) parameter boxes. In both the wide and narrow ranges, the emulator achieves percent-level accuracy with only mild exceptions at the lowest redshifts and smallest scales. We note that the overall mean error in the wide-prior range {\scriptsize Goku-W} suite, $0.45\%$, is substantially lower than its GP-based predecessor {\scriptsize GokuEmu}'s error of $2.92\%$, which used the same simulation data (see Fig.~13 of Ref.~\cite{Yang2025}). The worst-case error is improved even more significantly. For the {\scriptsize Goku-N} suite, the mean error is $0.18\%$, similar to the already small error of {\scriptsize GokuEmu} (see Fig.~14 of Ref.~\cite{Yang2025}). However, the error at small scales, especially at $k\gtrsim 2h/\text{Mpc}$ and $z=0$, is significantly reduced.

We tested the inference speed of {\scriptsize GokuNEmu} and found that it takes only $\sim 2\,$ms to make a prediction for a single cosmology on a laptop,\footnote{The reported runtime was measured on a MacBook Pro (M3). It may vary by a factor of a few on other machines.} two orders of magnitude faster than {\scriptsize EE2} ($\sim 300\,\text{ms}$ per evaluation)~\cite{Knabenhans2021} and about 7 times faster than the {\scriptsize CSST Emulator} ($\sim 15\,\text{ms}$ per evaluation)~\cite{Chen2025}.

\begin{figure*}
    \includegraphics[width=\linewidth]{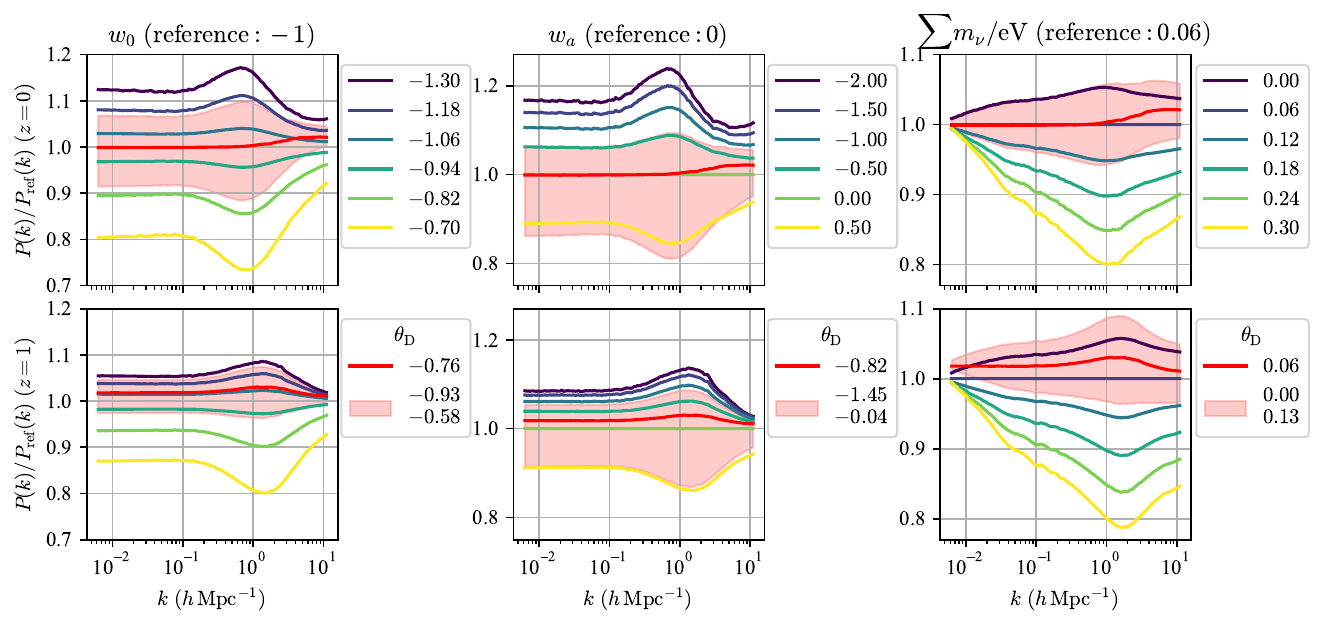}
    \caption{\label{fig:vary_param_DESI}Variation of the matter power spectrum induced by varying the cosmological parameters $w_0$, $w_a$, and $\sum m_\nu$ at $z=0$ and $1$. The curves are the ratio of the matter power spectrum at the varied parameter value to the spectrum at the reference cosmology ($\theta_\text{ref}$). The red lines show the ratio of the matter power spectrum of the DESI-like cosmology ($\theta_\text{D}$, with only $w_0$ and $w_a$ different from $\theta_\text{ref}$) to that of $\theta_\text{ref}$. The shaded regions indicate the corresponding 3$\sigma$ bounds.}
\end{figure*}
 
With the trained emulator, we perform a simple parameter sensitivity study by varying the parameters of dynamical DE and massive neutrinos, and measuring the change in the matter power spectrum at different redshifts. For a sensitivity study covering all cosmological parameters, see the supplemental material at the end of this document. The reference cosmology is defined as $\theta_\text{ref} = (\Omega_\text{m}, \Omega_\text{b}, h, A_\text{s},  n_\text{s}, w_0, w_a, \sum m_\nu, N_\text{eff}, \alpha_\text{s}) = (0.31, 0.048, 0.68, 2.1\times 10^{-9}, 0.96, -1, 0, 0.06\,\text{eV}, 3.044, 0)$. In addition, we define a cosmology similar to the constraint from DESI BAO+CMB+DESY5~\cite{Abdul-Karim2025} as $\theta_\text{D} = (w_0, w_a) = (-0.76, -0.82) $, with other parameters the same as the reference cosmology, and compare the matter power spectrum of $\theta_\text{D}$ to that of $\theta_\text{ref}$. Note that the DESI-like cosmology is not a perfect match to DESI, since the values of the other parameters are not guaranteed to be the same, but it is close enough to offer some insight into how a DESI preferred cosmology differs from a $\Lambda$CDM counterpart.

In Fig.~\ref{fig:vary_param_DESI}, we show the variation of the matter power spectrum induced by varying the cosmological parameters $w_0$, $w_a$, and $\sum m_\nu$. The first two columns show that the effects of time-dependent DE are more pronounced at lower redshifts, which is expected as the DE density becomes more significant compared to the matter density at lower redshifts. The third column shows that the effect of massive neutrinos is considerable at all redshifts. We see that the $k$ mode most sensitive to the variation of each of the parameters is $\sim 1h/\text{Mpc}$, deep in the nonlinear regime, for $ z < 1$. While $\theta_\text{D}$ is far away from the reference cosmology $\theta_\text{ref}$ in the $w_0$-$w_a$ plane, the matter power spectrum of $\theta_\text{D}$ (hereafter, $P_\text{D}$) differs from that of $\theta_\text{ref}$ by only a few percent. This is consistent with expectations, as the equation-of-state parameter of DE, $w(z)$, evolves from below $-1$ to above $-1$ over time (also known as ``phantom-crossing''), and the value never deviates too much from $-1$ at any redshift, with the given $w_0$ and $w_a$ values.\footnote{Indeed, $\theta_\mathrm{D}$ lies near the `mirage' line~\cite{Linder2008}, along which cosmologies yield similar expansion histories to that of $\Lambda$CDM.} We see that $P_\text{D}(z=0)$ is consistent with $P_\text{ref}(z=0)$ over large scales but deviates from it at small scales, implying information across scales can help break the degeneracy between the cosmological parameters. Also, we notice that $P_\text{D}(z=1)$ appears very similar to $P_\text{ref}^{w_0=-1.06} (z=1)$ (the cosmology differs from $\theta_\text{ref}$ only in $w_0$), while they are different at $z=0$, indicating that the redshift evolution of the matter power spectrum is useful for breaking degeneracies. Similar findings can also be observed for $P_\text{D}^{w_a = -0.04}$ and $P_\text{ref}^{w_a=0.5}$ (see the second column). We note the increase in $P_\text{D}(z=1)$ over $P_\text{ref}(z=1)$ mirrors the increase in the power spectrum for neutrino masses lower than $0.06\,$eV, the minimum from oscillation experiments. This illustrates one reason why a low $\sum m_\nu$ may be preferred in models where $w_0, w_a$ are enforced to be $(-1,0)$, away from their preferred values \cite{Elbers2025}.

\section{\label{sec:concl}Conclusion}

We have built the most powerful emulator for the nonlinear matter power spectrum to date, {\scriptsize GokuNEmu}, using the {\scriptsize T2N-MusE} framework. This emulator covers a 10D cosmological parameter space, including the base flat-$\Lambda$CDM parameters, the parameters for time-dependent DE, the sum of neutrino masses, the effective number of ultrarelativistic neutrinos, and the running of the primordial spectral index. It is capable of predicting the matter power spectrum at arbitrary redshifts between $0$ and $3$ and scales from $0.006h/\text{Mpc}$ to $10h/\text{Mpc}$, with percent-level accuracy (Fig.~\ref{fig:loo_WN}). It is also the fastest emulator for the matter power spectrum ($\sim 2\,$ms per evaluation).

Not only does {\scriptsize GokuNEmu} have an unprecedentedly high-dimensional parameter space, but it also has the widest prior coverage of the extended cosmological parameters. In particular, the prior coverage (Fig.~\ref{fig:prior_compare}) makes it the \textit{only} (besides its predecessor) matter power spectrum emulator that encompasses dynamical DE models from DESI~\cite{Abdul-Karim2025,Adame2025} and DES~\cite{Abbott2025} constraints. Current DESI DR1 results from the full-shape power spectrum~\cite{Adame2025b} improve the constraints only moderately from the BAO, due to the large number of free parameters required by the effective field theory nonlinear growth model, galaxy bias and instrumental models. Our approach allows a consistent modelling of structure growth to small, nonlinear scales, removing the need for some, though not all, of the modelling freedom and potentially improving constraints. In addition, the prior range of the sum of neutrino masses, $\sum m_\nu/\text{eV} \in [0, 0.6]$, covers recent constraints more safely.

{\scriptsize GokuNEmu} enables the breaking of parameter degeneracies by providing accurate predictions of structure growth across a range of redshifts and spatial scales (Fig.~\ref{fig:vary_param_DESI}). For example, we show an evolving DE model with parameters similar to those derived from DESI. Changes in $w_0$ and $w_a$ mirror the effect of low $\sum m_\nu < 0.06\,$eV. This degeneracy explains how parameter inference under a restricted $\Lambda$CDM cosmology, rather than $w_0 w_a$CDM, can drive an apparent preference for unphysically low neutrino masses \cite{Elbers2025}. The broad parameter coverage of our emulator makes it                                                                             particularly useful for cosmological analyses in future surveys that can measure cosmic structures with high precision down to nonlinear scales, such as LSST, \textit{Euclid}, the Roman Space Telescope, and CSST. We have made the emulator publicly available at Ref.~\cite{GokuNEmu2025}.

\section*{Acknowledgments}
We thank the anonymous referee and Ye (Issac) Lin for testing the emulator on different machines. We are grateful to Nicholas Kern for helpful discussions.
YY and SB acknowledge funding from NASA ATP 80NSSC22K1897. MFH is supported by the Leinweber Foundation and DOE grant DE-SC0019193. Computing resources were provided by Frontera LRAC AST21005.
The authors acknowledge the Frontera and Vista computing projects at the Texas Advanced Computing Center (TACC, \url{http://www.tacc.utexas.edu}) for providing HPC and storage resources that have contributed to the research results reported within this paper.
Frontera and Vista are made possible by National Science Foundation award OAC-1818253.

\bibliography{NEmu}

@Article{DESICollaboration2016,
       author = {{DESI Collaboration} and {Aghamousa}, Amir and {Aguilar et al.}, Jessica},
        title = "{The DESI Experiment Part I: Science,Targeting, and Survey Design}",
      journal = {arXiv e-prints},
         year = 2016,
        month = oct,
          eid = {arXiv:1611.00036},
        pages = {arXiv:1611.00036},
          doi = {10.48550/arXiv.1611.00036},
archivePrefix = {arXiv},
       eprint = {1611.00036},
 primaryClass = {astro-ph.IM},
       adsurl = {https://ui.adsabs.harvard.edu/abs/2016arXiv161100036D}
}

@Book{LSC2009,
  author        = {{Abell et al.}, Paul A.},
  title         = {LSST Science Book, Version 2.0},
  year          = {2009},
  month         = dec,
  archiveprefix = {arXiv},
  copyright     = {arXiv.org perpetual, non-exclusive license},
  doi           = {10.48550/ARXIV.0912.0201},
  eprint        = {0912.0201},
  primaryclass  = {astro-ph.IM},
  publisher     = {arXiv},
}

@Article{Laureijs2011,
  author        = {{Laureijs et al.}, R.},
  journal       = {arXiv e-prints},
  title         = "{Euclid Definition Study Report}",
  year          = {2011},
  month         = oct,
  pages         = {arXiv:1110.3193},
  archiveprefix = {arXiv},
  doi           = {10.48550/arXiv.1110.3193},
  eid           = {arXiv:1110.3193},
  eprint        = {1110.3193},
  primaryclass  = {astro-ph.CO},
  url           = {https://ui.adsabs.harvard.edu/abs/2011arXiv1110.3193L},
}

@Article{Akeson2019,
       author = {{Akeson et al.}, Rachel},
        title = "{The Wide Field Infrared Survey Telescope: 100 Hubbles for the 2020s}",
      journal = {arXiv e-prints},
         year = 2019,
        month = feb,
          eid = {arXiv:1902.05569},
        pages = {arXiv:1902.05569},
          doi = {10.48550/arXiv.1902.05569},
archivePrefix = {arXiv},
       eprint = {1902.05569},
 primaryClass = {astro-ph.IM},
       adsurl = {https://ui.adsabs.harvard.edu/abs/2019arXiv190205569A}
}

@Article{Heitmann2009,
  author    = {Heitmann, Katrin and Higdon, David and White, Martin and Habib, Salman and Williams, Brian J. and Lawrence, Earl and Wagner, Christian},
  journal   = {The Astrophysical Journal},
  title     = "{THE COYOTE UNIVERSE. II. COSMOLOGICAL MODELS AND PRECISION EMULATION OF THE NONLINEAR MATTER POWER SPECTRUM}",
  year      = {2009},
  issn      = {1538-4357},
  month     = oct,
  number    = {1},
  pages     = {156--174},
  volume    = {705},
  doi       = {10.1088/0004-637x/705/1/156},
  publisher = {American Astronomical Society},
}

@Article{Heitmann2010,
  author    = {Heitmann, Katrin and White, Martin and Wagner, Christian and Habib, Salman and Higdon, David},
  journal   = {The Astrophysical Journal},
  title     = "{THE COYOTE UNIVERSE. I. PRECISION DETERMINATION OF THE NONLINEAR MATTER POWER SPECTRUM}",
  year      = {2010},
  issn      = {1538-4357},
  month     = apr,
  number    = {1},
  pages     = {104--121},
  volume    = {715},
  doi       = {10.1088/0004-637x/715/1/104},
  publisher = {American Astronomical Society},
}

@Article{Heitmann2013,
  author    = {Heitmann, Katrin and Lawrence, Earl and Kwan, Juliana and Habib, Salman and Higdon, David},
  journal   = {The Astrophysical Journal},
  title     = "{THE COYOTE UNIVERSE EXTENDED: PRECISION EMULATION OF THE MATTER POWER SPECTRUM}",
  year      = {2013},
  issn      = {1538-4357},
  month     = dec,
  number    = {1},
  pages     = {111},
  volume    = {780},
  doi       = {10.1088/0004-637x/780/1/111},
  publisher = {American Astronomical Society},
}

@Article{Heitmann2016,
  author    = {Heitmann, Katrin and Bingham, Derek and Lawrence, Earl and Bergner, Steven and Habib, Salman and Higdon, David and Pope, Adrian and Biswas, Rahul and Finkel, Hal and Frontiere, Nicholas and Bhattacharya, Suman},
  journal   = {The Astrophysical Journal},
  title     = "{THE MIRA–TITAN UNIVERSE: PRECISION PREDICTIONS FOR DARK ENERGY SURVEYS}",
  year      = {2016},
  issn      = {1538-4357},
  month     = mar,
  number    = {2},
  pages     = {108},
  volume    = {820},
  doi       = {10.3847/0004-637x/820/2/108},
  publisher = {American Astronomical Society},
}

@Article{Lawrence2017,
  author    = {Lawrence, Earl and Heitmann, Katrin and Kwan, Juliana and Upadhye, Amol and Bingham, Derek and Habib, Salman and Higdon, David and Pope, Adrian and Finkel, Hal and Frontiere, Nicholas},
  journal   = {The Astrophysical Journal},
  title     = "{The Mira-Titan Universe. II. Matter Power Spectrum Emulation}",
  year      = {2017},
  issn      = {1538-4357},
  month     = sep,
  number    = {1},
  pages     = {50},
  volume    = {847},
  doi       = {10.3847/1538-4357/aa86a9},
  publisher = {American Astronomical Society},
}

@Article{DeRose2019,
  author    = {DeRose, Joseph and Wechsler, Risa H. and Tinker, Jeremy L. and Becker, Matthew R. and Mao, Yao-Yuan and McClintock, Thomas and McLaughlin, Sean and Rozo, Eduardo and Zhai, Zhongxu},
  journal   = {The Astrophysical Journal},
  title     = "{The Aemulus Project. I. Numerical Simulations for Precision Cosmology}",
  year      = {2019},
  issn      = {1538-4357},
  month     = apr,
  number    = {1},
  pages     = {69},
  volume    = {875},
  doi       = {10.3847/1538-4357/ab1085},
  publisher = {American Astronomical Society},
}

@Article{McClintock2019,
  author    = {McClintock, Thomas and Rozo, Eduardo and Becker, Matthew R. and DeRose, Joseph and Mao, Yao-Yuan and McLaughlin, Sean and Tinker, Jeremy L. and Wechsler, Risa H. and Zhai, Zhongxu},
  journal   = {The Astrophysical Journal},
  title     = "{The Aemulus Project. II. Emulating the Halo Mass Function}",
  year      = {2019},
  issn      = {1538-4357},
  month     = feb,
  number    = {1},
  pages     = {53},
  volume    = {872},
  doi       = {10.3847/1538-4357/aaf568},
  publisher = {American Astronomical Society},
}

@Article{Zhai2019,
  author    = {Zhai, Zhongxu and Tinker, Jeremy L. and Becker, Matthew R. and DeRose, Joseph and Mao, Yao-Yuan and McClintock, Thomas and McLaughlin, Sean and Rozo, Eduardo and Wechsler, Risa H.},
  journal   = {The Astrophysical Journal},
  title     = "{The Aemulus Project. III. Emulation of the Galaxy Correlation Function}",
  year      = {2019},
  issn      = {1538-4357},
  month     = mar,
  number    = {1},
  pages     = {95},
  volume    = {874},
  doi       = {10.3847/1538-4357/ab0d7b},
  publisher = {American Astronomical Society},
}

@Article{Smith2019,
  author    = {Smith, Robert E and Angulo, Raul E},
  journal   = {Monthly Notices of the Royal Astronomical Society},
  title     = "{Precision modelling of the matter power spectrum in a Planck-like Universe}",
  year      = {2019},
  issn      = {1365-2966},
  month     = apr,
  number    = {1},
  pages     = {1448--1479},
  volume    = {486},
  doi       = {10.1093/mnras/stz890},
  publisher = {Oxford University Press (OUP)},
}

@article{Knabenhans2019,
    author = {{Euclid Collaboration} and {Knabenhans}, Mischa and {Stadel et al.}, Joachim},
    title = "{Euclid preparation: II. The EuclidEmulator – a tool to compute the cosmology dependence of the nonlinear matter power spectrum}",
    journal = {Monthly Notices of the Royal Astronomical Society},
    volume = {484},
    number = {4},
    pages = {5509-5529},
    year = {2019},
    month = {01},
    issn = {0035-8711},
    doi = {10.1093/mnras/stz197},
    url = {https://doi.org/10.1093/mnras/stz197},
    eprint = {https://academic.oup.com/mnras/article-pdf/484/4/5509/27790453/stz197.pdf},
}

@Article{Knabenhans2021,
  author    = {{Euclid Collaboration} and {Knabenhans}, M. and {Stadel et al.}, J.},
  journal   = {Monthly Notices of the Royal Astronomical Society},
  title     = "{Euclid preparation: IX. EuclidEmulator2 – power spectrum emulation with massive neutrinos and self-consistent dark energy perturbations}",
  year      = {2021},
  issn      = {1365-2966},
  month     = may,
  number    = {2},
  pages     = {2840--2869},
  volume    = {505},
  doi       = {10.1093/mnras/stab1366},
  publisher = {Oxford University Press (OUP)},
}

@Article{Valcin2019,
  author    = {Valcin, David and Villaescusa-Navarro, Francisco and Verde, Licia and Raccanelli, Alvise},
  journal   = {Journal of Cosmology and Astroparticle Physics},
  title     = "{BE-HaPPY: bias emulator for halo power spectrum including massive neutrinos}",
  year      = {2019},
  issn      = {1475-7516},
  month     = dec,
  number    = {12},
  pages     = {057--057},
  volume    = {2019},
  doi       = {10.1088/1475-7516/2019/12/057},
  publisher = {IOP Publishing},
}

@Article{Nishimichi2019,
  author    = {Nishimichi, Takahiro and Takada, Masahiro and Takahashi, Ryuichi and Osato, Ken and Shirasaki, Masato and Oogi, Taira and Miyatake, Hironao and Oguri, Masamune and Murata, Ryoma and Kobayashi, Yosuke and Yoshida, Naoki},
  journal   = {The Astrophysical Journal},
  title     = "{Dark Quest. I. Fast and Accurate Emulation of Halo Clustering Statistics and Its Application to Galaxy Clustering}",
  year      = {2019},
  issn      = {1538-4357},
  month     = oct,
  number    = {1},
  pages     = {29},
  volume    = {884},
  doi       = {10.3847/1538-4357/ab3719},
  publisher = {American Astronomical Society},
}

@Article{Caldwell2009,
       author = {{Caldwell}, Robert R. and {Kamionkowski}, Marc},
        title = "{The Physics of Cosmic Acceleration}",
      journal = {Annual Review of Nuclear and Particle Science},
         year = 2009,
        month = nov,
       volume = {59},
       number = {1},
        pages = {397-429},
          doi = {10.1146/annurev-nucl-010709-151330},
archivePrefix = {arXiv},
       eprint = {0903.0866},
 primaryClass = {astro-ph.CO},
       adsurl = {https://ui.adsabs.harvard.edu/abs/2009ARNPS..59..397C}
}

@Article{Feng2010,
       author = {{Feng}, Jonathan L.},
        title = "{Dark Matter Candidates from Particle Physics and Methods of Detection}",
      journal = {Annual Review of Astronomy and Astrophysics},
         year = 2010,
        month = sep,
       volume = {48},
        pages = {495-545},
          doi = {10.1146/annurev-astro-082708-101659},
archivePrefix = {arXiv},
       eprint = {1003.0904},
 primaryClass = {astro-ph.CO},
       adsurl = {https://ui.adsabs.harvard.edu/abs/2010ARA&A..48..495F}
}

@Article{Wong2011,
       author = {{Wong}, Yvonne Y.~Y.},
        title = "{Neutrino Mass in Cosmology: Status and Prospects}",
      journal = {Annual Review of Nuclear and Particle Science},
         year = 2011,
        month = nov,
       volume = {61},
       number = {1},
        pages = {69-98},
          doi = {10.1146/annurev-nucl-102010-130252},
archivePrefix = {arXiv},
       eprint = {1111.1436},
 primaryClass = {astro-ph.CO},
       adsurl = {https://ui.adsabs.harvard.edu/abs/2011ARNPS..61...69W}
}

@ARTICLE{Ho2023,
       author = {{Ho}, Ming-Feng and {Bird}, Simeon and {Fernandez}, Martin A. and {Shelton}, Christian R.},
        title = "{MF-Box: multifidelity and multiscale emulation for the matter power spectrum}",
      journal = {MNRAS},
         year = 2023,
        month = dec,
       volume = {526},
       number = {2},
        pages = {2903-2919},
          doi = {10.1093/mnras/stad2901},
archivePrefix = {arXiv},
       eprint = {2306.03144},
 primaryClass = {astro-ph.CO},
       adsurl = {https://ui.adsabs.harvard.edu/abs/2023MNRAS.526.2903H}
}

@ARTICLE{Ho2022,
       author = {{Ho}, Ming-Feng and {Bird}, Simeon and {Shelton}, Christian R.},
        title = "{Multifidelity emulation for the matter power spectrum using Gaussian processes}",
      journal = {MNRAS},
         year = 2022,
        month = jan,
       volume = {509},
       number = {2},
        pages = {2551-2565},
          doi = {10.1093/mnras/stab3114},
archivePrefix = {arXiv},
       eprint = {2105.01081},
 primaryClass = {astro-ph.CO},
       adsurl = {https://ui.adsabs.harvard.edu/abs/2022MNRAS.509.2551H}
}

@ARTICLE{Ali2013,
       author = {{Ali-Ha{\"\i}moud}, Yacine and {Bird}, Simeon},
        title = "{An efficient implementation of massive neutrinos in non-linear structure formation simulations}",
      journal = {MNRAS},
         year = 2013,
        month = feb,
       volume = {428},
       number = {4},
        pages = {3375-3389},
          doi = {10.1093/mnras/sts286},
archivePrefix = {arXiv},
       eprint = {1209.0461},
 primaryClass = {astro-ph.CO},
       adsurl = {https://ui.adsabs.harvard.edu/abs/2013MNRAS.428.3375A}
}

@ARTICLE{Bird2018,
       author = {{Bird}, Simeon and {Ali-Ha{\"\i}moud}, Yacine and {Feng}, Yu and {Liu}, Jia},
        title = "{An efficient and accurate hybrid method for simulating non-linear neutrino structure}",
      journal = {MNRAS},
         year = 2018,
        month = dec,
       volume = {481},
       number = {2},
        pages = {1486-1500},
          doi = {10.1093/mnras/sty2376},
archivePrefix = {arXiv},
       eprint = {1803.09854},
 primaryClass = {astro-ph.CO},
       adsurl = {https://ui.adsabs.harvard.edu/abs/2018MNRAS.481.1486B}
}

@software{Feng2018,
  author       = {Yu Feng and
                  Simeon Bird and
                  Lauren Anderson and
                  Andreu Font-Ribera and
                  Chris Pedersen},
  title        = "{MP-Gadget/MP-Gadget: A tag for getting a DOI}",
  month        = oct,
  year         = 2018,
  publisher    = {Zenodo},
  version      = {FirstDOI},
  doi          = {10.5281/zenodo.1451799},
  url          = {https://doi.org/10.5281/zenodo.1451799}
}

@article{Ba2015,
author = {Shan Ba, William R. Myers and William A. Brenneman},
title = "{Optimal Sliced Latin Hypercube Designs}",
journal = {Technometrics},
volume = {57},
number = {4},
pages = {479-487},
year = {2015},
publisher = {Taylor & Francis},
doi = {10.1080/00401706.2014.957867},


URL = { 
    
        https://doi.org/10.1080/00401706.2014.957867
    
    

},
eprint = { 
    
        https://doi.org/10.1080/00401706.2014.957867
    
    

}

}

@ARTICLE{Aghanim2020,
       author = {{Planck Collaboration} and {Aghanim}, N. and {Akrami et al.}, Y.},
        title = "{Planck 2018 results. VI. Cosmological parameters}",
      journal = {\aap},
         year = 2020,
        month = sep,
       volume = {641},
          eid = {A6},
        pages = {A6},
          doi = {10.1051/0004-6361/201833910},
archivePrefix = {arXiv},
       eprint = {1807.06209},
 primaryClass = {astro-ph.CO},
       adsurl = {https://ui.adsabs.harvard.edu/abs/2020A&A...641A...6P}
}

@ARTICLE{Riess2022,
       author = {{Riess et al.}, Adam G.},
        title = "{A Comprehensive Measurement of the Local Value of the Hubble Constant with 1 km s$^{-1}$ Mpc$^{-1}$ Uncertainty from the Hubble Space Telescope and the SH0ES Team}",
      journal = {\apjl},
         year = 2022,
        month = jul,
       volume = {934},
       number = {1},
          eid = {L7},
        pages = {L7},
          doi = {10.3847/2041-8213/ac5c5b},
archivePrefix = {arXiv},
       eprint = {2112.04510},
 primaryClass = {astro-ph.CO},
       adsurl = {https://ui.adsabs.harvard.edu/abs/2022ApJ...934L...7R}
}

@article{Shvartsman1969,
    author = "Shvartsman, V. F.",
    title = "{Density of relict particles with zero rest mass in the universe}",
    journal = "Pisma Zh. Eksp. Teor. Fiz.",
    volume = "9",
    pages = "315--317",
    year = "1969"
}

@article{Steigman1977,
    author = "Steigman, G. and Schramm, D. N. and Gunn, J. E.",
    title = "{Cosmological Limits to the Number of Massive Leptons}",
    doi = "10.1016/0370-2693(77)90176-9",
    journal = "Phys. Lett. B",
    volume = "66",
    pages = "202--204",
    year = "1977"
}

@ARTICLE{Lesgourgues2006,
       author = {{Lesgourgues}, Julien and {Pastor}, Sergio},
        title = "{Massive neutrinos and cosmology}",
      journal = {\physrep},
         year = 2006,
        month = jul,
       volume = {429},
       number = {6},
        pages = {307-379},
          doi = {10.1016/j.physrep.2006.04.001},
archivePrefix = {arXiv},
       eprint = {astro-ph/0603494},
 primaryClass = {astro-ph},
       adsurl = {https://ui.adsabs.harvard.edu/abs/2006PhR...429..307L}
}

@ARTICLE{Angulo2016,
       author = {{Angulo}, Raul E. and {Pontzen}, Andrew},
        title = "{Cosmological N-body simulations with suppressed variance}",
      journal = {\mnras},
         year = 2016,
        month = oct,
       volume = {462},
       number = {1},
        pages = {L1-L5},
          doi = {10.1093/mnrasl/slw098},
archivePrefix = {arXiv},
       eprint = {1603.05253},
 primaryClass = {astro-ph.CO},
       adsurl = {https://ui.adsabs.harvard.edu/abs/2016MNRAS.462L...1A}
}

@ARTICLE{Arico2021,
       author = {{Aric{\`o}}, Giovanni and {Angulo}, Raul E. and {Contreras}, Sergio and {Ondaro-Mallea}, Lurdes and {Pellejero-Iba{\~n}ez}, Marcos and {Zennaro}, Matteo},
        title = "{The BACCO simulation project: a baryonification emulator with neural networks}",
      journal = {\mnras},
         year = 2021,
        month = sep,
       volume = {506},
       number = {3},
        pages = {4070-4082},
          doi = {10.1093/mnras/stab1911},
archivePrefix = {arXiv},
       eprint = {2011.15018},
 primaryClass = {astro-ph.CO},
       adsurl = {https://ui.adsabs.harvard.edu/abs/2021MNRAS.506.4070A}
}

@ARTICLE{Villaescusa2020,
       author = {{Villaescusa-Navarro}, Francisco and {Hahn}, ChangHoon and {Massara}, Elena and {Banerjee}, Arka and {Delgado}, Ana Maria and {Ramanah}, Doogesh Kodi and {Charnock}, Tom and {Giusarma}, Elena and {Li}, Yin and {Allys}, Erwan and {Brochard}, Antoine and {Uhlemann}, Cora and {Chiang}, Chi-Ting and {He}, Siyu and {Pisani}, Alice and {Obuljen}, Andrej and {Feng}, Yu and {Castorina}, Emanuele and {Contardo}, Gabriella and {Kreisch}, Christina D. and {Nicola}, Andrina and {Alsing}, Justin and {Scoccimarro}, Roman and {Verde}, Licia and {Viel}, Matteo and {Ho}, Shirley and {Mallat}, Stephane and {Wandelt}, Benjamin and {Spergel}, David N.},
        title = "{The Quijote Simulations}",
      journal = {\apjs},
         year = 2020,
        month = sep,
       volume = {250},
       number = {1},
          eid = {2},
        pages = {2},
          doi = {10.3847/1538-4365/ab9d82},
archivePrefix = {arXiv},
       eprint = {1909.05273},
 primaryClass = {astro-ph.CO},
       adsurl = {https://ui.adsabs.harvard.edu/abs/2020ApJS..250....2V}
}

@ARTICLE{Riess2021,
       author = {{Riess}, Adam G. and {Casertano}, Stefano and {Yuan}, Wenlong and {Bowers}, J. Bradley and {Macri}, Lucas and {Zinn}, Joel C. and {Scolnic}, Dan},
        title = "{Cosmic Distances Calibrated to 1\% Precision with Gaia EDR3 Parallaxes and Hubble Space Telescope Photometry of 75 Milky Way Cepheids Confirm Tension with {\ensuremath{\Lambda}}CDM}",
      journal = {\apjl},
         year = 2021,
        month = feb,
       volume = {908},
       number = {1},
          eid = {L6},
        pages = {L6},
          doi = {10.3847/2041-8213/abdbaf},
archivePrefix = {arXiv},
       eprint = {2012.08534},
 primaryClass = {astro-ph.CO},
       adsurl = {https://ui.adsabs.harvard.edu/abs/2021ApJ...908L...6R}
}

@ARTICLE{Abbott2022,
       author = {{DES Collaboration} and {Abbott}, T.~M.~C. and {Aguena et al.}, M.},
        title = "{Dark Energy Survey Year 3 results: Cosmological constraints from galaxy clustering and weak lensing}",
      journal = {\prd},
         year = 2022,
        month = jan,
       volume = {105},
       number = {2},
          eid = {023520},
        pages = {023520},
          doi = {10.1103/PhysRevD.105.023520},
archivePrefix = {arXiv},
       eprint = {2105.13549},
 primaryClass = {astro-ph.CO},
       adsurl = {https://ui.adsabs.harvard.edu/abs/2022PhRvD.105b3520A}
}

@ARTICLE{Asgari2021,
       author = {{Asgari}, Marika and {Lin}, Chieh-An and {Joachimi}, Benjamin and {Giblin}, Benjamin and {Heymans}, Catherine and {Hildebrandt}, Hendrik and {Kannawadi}, Arun and {St{\"o}lzner}, Benjamin and {Tr{\"o}ster}, Tilman and {van den Busch}, Jan Luca and {Wright}, Angus H. and {Bilicki}, Maciej and {Blake}, Chris and {de Jong}, Jelte and {Dvornik}, Andrej and {Erben}, Thomas and {Getman}, Fedor and {Hoekstra}, Henk and {K{\"o}hlinger}, Fabian and {Kuijken}, Konrad and {Miller}, Lance and {Radovich}, Mario and {Schneider}, Peter and {Shan}, HuanYuan and {Valentijn}, Edwin},
        title = "{KiDS-1000 cosmology: Cosmic shear constraints and comparison between two point statistics}",
      journal = {\aap},
         year = 2021,
        month = jan,
       volume = {645},
          eid = {A104},
        pages = {A104},
          doi = {10.1051/0004-6361/202039070},
archivePrefix = {arXiv},
       eprint = {2007.15633},
 primaryClass = {astro-ph.CO},
       adsurl = {https://ui.adsabs.harvard.edu/abs/2021A&A...645A.104A}
}

@ARTICLE{Linder2003,
       author = {{Linder}, Eric V.},
        title = "{Exploring the Expansion History of the Universe}",
      journal = {\prl},
         year = 2003,
        month = mar,
       volume = {90},
       number = {9},
          eid = {091301},
        pages = {091301},
          doi = {10.1103/PhysRevLett.90.091301},
archivePrefix = {arXiv},
       eprint = {astro-ph/0208512},
 primaryClass = {astro-ph},
       adsurl = {https://ui.adsabs.harvard.edu/abs/2003PhRvL..90i1301L}
}

@ARTICLE{Bocquet2020,
       author = {{Bocquet}, Sebastian and {Heitmann}, Katrin and {Habib}, Salman and {Lawrence}, Earl and {Uram}, Thomas and {Frontiere}, Nicholas and {Pope}, Adrian and {Finkel}, Hal},
        title = "{The Mira-Titan Universe. III. Emulation of the Halo Mass Function}",
      journal = {\apj},
         year = 2020,
        month = sep,
       volume = {901},
       number = {1},
          eid = {5},
        pages = {5},
          doi = {10.3847/1538-4357/abac5c},
archivePrefix = {arXiv},
       eprint = {2003.12116},
 primaryClass = {astro-ph.CO},
       adsurl = {https://ui.adsabs.harvard.edu/abs/2020ApJ...901....5B}
}

@ARTICLE{Kwan2023,
       author = {{Kwan}, Juliana and {Saito}, Shun and {Leauthaud}, Alexie and {Heitmann}, Katrin and {Habib}, Salman and {Frontiere}, Nicholas and {Guo}, Hong and {Huang}, Song and {Pope}, Adrian and {Rodrigu{\'e}z-Torres}, Sergio},
        title = "{Galaxy Clustering in the Mira-Titan Universe. I. Emulators for the Redshift Space Galaxy Correlation Function and Galaxy-Galaxy Lensing}",
      journal = {\apj},
         year = 2023,
        month = jul,
       volume = {952},
       number = {1},
          eid = {80},
        pages = {80},
          doi = {10.3847/1538-4357/acd92f},
archivePrefix = {arXiv},
       eprint = {2302.12379},
 primaryClass = {astro-ph.CO},
       adsurl = {https://ui.adsabs.harvard.edu/abs/2023ApJ...952...80K}
}

@ARTICLE{Moran2023,
       author = {{Moran}, Kelly R. and {Heitmann}, Katrin and {Lawrence}, Earl and {Habib}, Salman and {Bingham}, Derek and {Upadhye}, Amol and {Kwan}, Juliana and {Higdon}, David and {Payne}, Richard},
        title = "{The Mira-Titan Universe - IV. High-precision power spectrum emulation}",
      journal = {\mnras},
         year = 2023,
        month = apr,
       volume = {520},
       number = {3},
        pages = {3443-3458},
          doi = {10.1093/mnras/stac3452},
archivePrefix = {arXiv},
       eprint = {2207.12345},
 primaryClass = {astro-ph.CO},
       adsurl = {https://ui.adsabs.harvard.edu/abs/2023MNRAS.520.3443M}
}

@ARTICLE{Chevallier2001,
       author = {{Chevallier}, Michel and {Polarski}, David},
        title = "{Accelerating Universes with Scaling Dark Matter}",
      journal = {International Journal of Modern Physics D},
         year = 2001,
        month = jan,
       volume = {10},
       number = {2},
        pages = {213-223},
          doi = {10.1142/S0218271801000822},
archivePrefix = {arXiv},
       eprint = {gr-qc/0009008},
 primaryClass = {gr-qc},
       adsurl = {https://ui.adsabs.harvard.edu/abs/2001IJMPD..10..213C}
}

@ARTICLE{Takada2014,
       author = {{Takada}, Masahiro and {Ellis}, Richard S. and {Chiba}, Masashi and {Greene}, Jenny E. and {Aihara}, Hiroaki and {Arimoto}, Nobuo and {Bundy}, Kevin and {Cohen}, Judith and {Dor{\'e}}, Olivier and {Graves}, Genevieve and {Gunn}, James E. and {Heckman}, Timothy and {Hirata}, Christopher M. and {Ho}, Paul and {Kneib}, Jean-Paul and {Le F{\`e}vre}, Olivier and {Lin}, Lihwai and {More}, Surhud and {Murayama}, Hitoshi and {Nagao}, Tohru and {Ouchi}, Masami and {Seiffert}, Michael and {Silverman}, John D. and {Sodr{\'e}}, Laerte and {Spergel}, David N. and {Strauss}, Michael A. and {Sugai}, Hajime and {Suto}, Yasushi and {Takami}, Hideki and {Wyse}, Rosemary},
        title = "{Extragalactic science, cosmology, and Galactic archaeology with the Subaru Prime Focus Spectrograph}",
      journal = {\pasj},
         year = 2014,
        month = feb,
       volume = {66},
       number = {1},
          eid = {R1},
        pages = {R1},
          doi = {10.1093/pasj/pst019},
archivePrefix = {arXiv},
       eprint = {1206.0737},
 primaryClass = {astro-ph.CO},
       adsurl = {https://ui.adsabs.harvard.edu/abs/2014PASJ...66R...1T}
}

@ARTICLE{Casares2024,
       author = {{S{\'a}ez-Casares}, I. and {Rasera}, Y. and {Richardson}, T.~R.~G. and {Corasaniti}, P. -S.},
        title = "{The e-MANTIS emulator: Fast and accurate predictions of the halo mass function in f(R)CDM and wCDM cosmologies}",
      journal = {\aap},
         year = 2024,
        month = nov,
       volume = {691},
          eid = {A323},
        pages = {A323},
          doi = {10.1051/0004-6361/202450193},
archivePrefix = {arXiv},
       eprint = {2410.05226},
 primaryClass = {astro-ph.CO},
       adsurl = {https://ui.adsabs.harvard.edu/abs/2024A&A...691A.323S}
}

@ARTICLE{Chen2025,
       author = {{Chen}, Zhao and {Yu}, Yu and {Han}, Jiaxin and {Jing}, Y.~P.},
        title = "{CSST Cosmological Emulator I: Matter Power Spectrum Emulation with one percent accuracy}",
      journal = {arXiv e-prints},
         year = 2025,
        month = feb,
          eid = {arXiv:2502.11160},
        pages = {arXiv:2502.11160},
          doi = {10.48550/arXiv.2502.11160},
archivePrefix = {arXiv},
       eprint = {2502.11160},
 primaryClass = {astro-ph.CO},
       adsurl = {https://ui.adsabs.harvard.edu/abs/2025arXiv250211160C}
}

@ARTICLE{Gong2019,
       author = {{Gong}, Yan and {Liu}, Xiangkun and {Cao}, Ye and {Chen}, Xuelei and {Fan}, Zuhui and {Li}, Ran and {Li}, Xiao-Dong and {Li}, Zhigang and {Zhang}, Xin and {Zhan}, Hu},
        title = "{Cosmology from the Chinese Space Station Optical Survey (CSS-OS)}",
      journal = {\apj},
         year = 2019,
        month = oct,
       volume = {883},
       number = {2},
          eid = {203},
        pages = {203},
          doi = {10.3847/1538-4357/ab391e},
archivePrefix = {arXiv},
       eprint = {1901.04634},
 primaryClass = {astro-ph.CO},
       adsurl = {https://ui.adsabs.harvard.edu/abs/2019ApJ...883..203G}
}

@ARTICLE{Yang2025,
       author = {{Yang}, Yanhui and {Bird}, Simeon and {Ho}, Ming-Feng},
        title = "{Ten-parameter simulation suite for cosmological emulation beyond {\ensuremath{\Lambda}}CDM}",
      journal = {\prd},
         year = 2025,
        month = apr,
       volume = {111},
       number = {8},
          eid = {083529},
        pages = {083529},
          doi = {10.1103/PhysRevD.111.083529},
archivePrefix = {arXiv},
       eprint = {2501.06296},
 primaryClass = {astro-ph.CO},
       adsurl = {https://ui.adsabs.harvard.edu/abs/2025PhRvD.111h3529Y}
}

@ARTICLE{Yang2025b,
       author = {{Yang}, Yanhui and {Bird}, Simeon and {Ho}, Ming-Feng and {Qezlou}, Mahdi},
        title = "{Design and optimization of neural networks for multifidelity cosmological emulation}",
      journal = {arXiv e-prints},
         year = 2025,
        month = jul,
          eid = {arXiv:2507.07184},
        pages = {arXiv:2507.07184},
          doi = {10.48550/arXiv.2507.07184},
archivePrefix = {arXiv},
       eprint = {2507.07184},
 primaryClass = {astro-ph.CO},
       adsurl = {https://ui.adsabs.harvard.edu/abs/2025arXiv250707184Y}
}

@ARTICLE{Cabayol-Garcia2023,
       author = {{Cabayol-Garcia}, L. and {Chaves-Montero}, J. and {Font-Ribera}, A. and {Pedersen}, C.},
        title = "{A neural network emulator for the Lyman-{\ensuremath{\alpha}} forest 1D flux power spectrum}",
      journal = {\mnras},
         year = 2023,
        month = nov,
       volume = {525},
       number = {3},
        pages = {3499-3515},
          doi = {10.1093/mnras/stad2512},
archivePrefix = {arXiv},
       eprint = {2305.19064},
 primaryClass = {astro-ph.CO},
       adsurl = {https://ui.adsabs.harvard.edu/abs/2023MNRAS.525.3499C}
}

@ARTICLE{Diao2025,
       author = {{Diao}, Kangning and {Mao}, Yi},
        title = "{Multi-fidelity emulator for large-scale 21 cm lightcone images: a few-shot transfer learning approach with generative adversarial network}",
      journal = {arXiv e-prints},
         year = 2025,
        month = feb,
          eid = {arXiv:2502.04246},
        pages = {arXiv:2502.04246},
          doi = {10.48550/arXiv.2502.04246},
archivePrefix = {arXiv},
       eprint = {2502.04246},
 primaryClass = {astro-ph.IM},
       adsurl = {https://ui.adsabs.harvard.edu/abs/2025arXiv250204246D}
}

@ARTICLE{Abdul-Karim2025,
       author = {{DESI Collaboration} and {Abdul-Karim}, M. and {Aguilar et al.}, J.},
        title = "{DESI DR2 Results II: Measurements of Baryon Acoustic Oscillations and Cosmological Constraints}",
      journal = {arXiv e-prints},
         year = 2025,
        month = mar,
          eid = {arXiv:2503.14738},
        pages = {arXiv:2503.14738},
          doi = {10.48550/arXiv.2503.14738},
archivePrefix = {arXiv},
       eprint = {2503.14738},
 primaryClass = {astro-ph.CO},
       adsurl = {https://ui.adsabs.harvard.edu/abs/2025arXiv250314738D}
}

@misc{GokuNEmu2025,
  author       = {{Yang}, Yanhui and {Bird}, Simeon and {Ho}, Ming-Feng and {Qezlou}, Mahdi},
  title        = "{GokuNEmu: A Neural Network Emulator Based on the Goku Simulation Suite}",
  year         = {2025},
  url          = {https://github.com/astro-YYH/GokuNEmu},
  note         = "{GitHub repository}"
}

\appendix
\section*{\label{app:param_sensitivity}Parameter Sensitivity}

\begin{figure*}
    \centering
    \includegraphics[width=\linewidth]{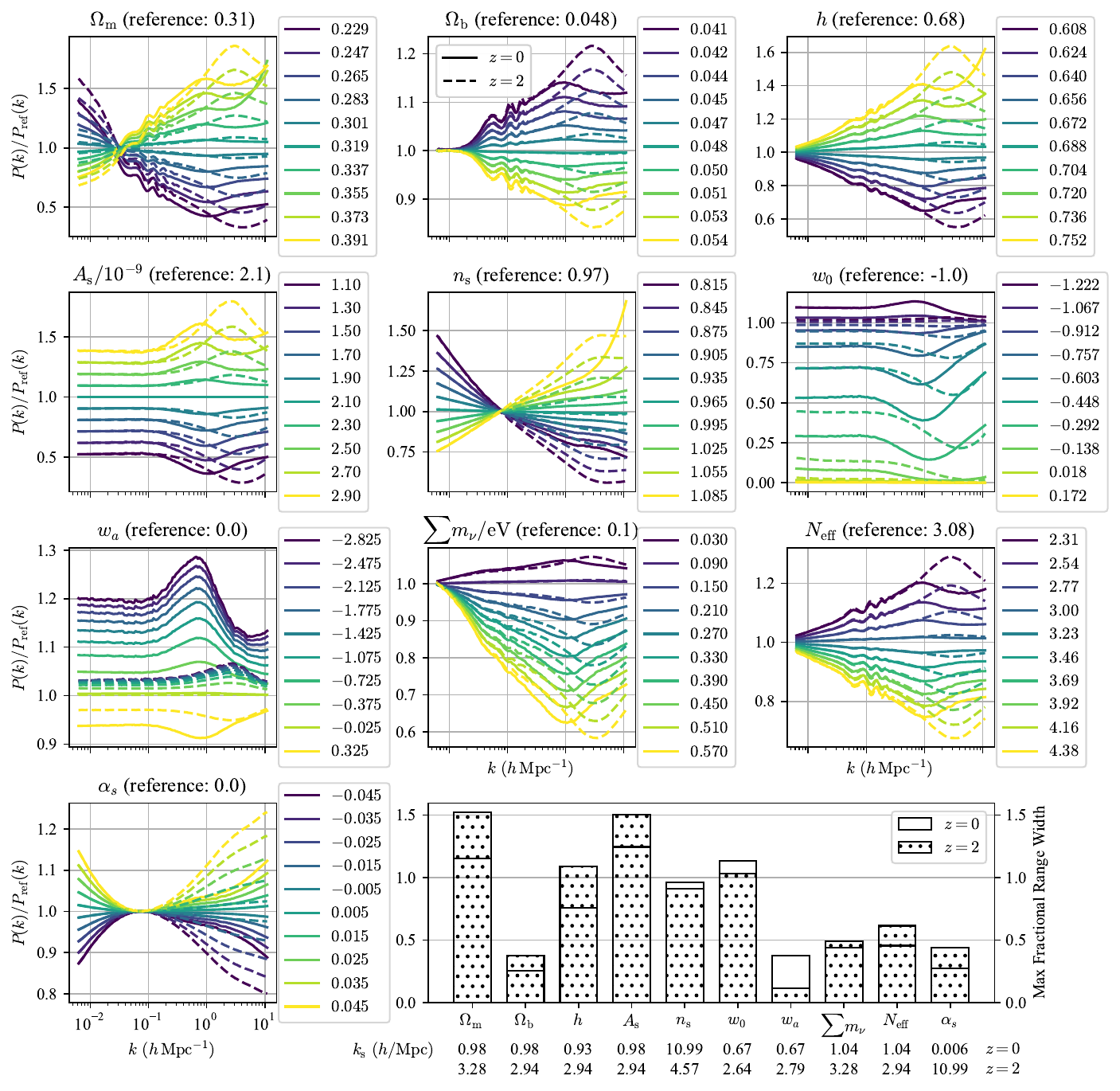}
    \caption{\label{fig:param_sensitivity} Panels 1 to 10: variations in the matter power spectrum at $z=0$ induced by each of 10 cosmological parameters, evaluated at the reference cosmology (defined in Ref.~\cite{Yang2025}). In each panel, one parameter is varied across 10 values, spanning from 5\% to 95\% of its prior range, while all other parameters are fixed at their reference values. The power spectra are normalized to the reference spectrum, with distinct colors indicating different parameter values. When we refer to panels 1 to 10, they are arranged sequentially from left to right and then top to bottom. Panel 11: the size of the fractional range of variation (difference between the largest and smallest power values across $k$ modes) of the matter power spectrum, relative to the reference cosmology, is shown for each parameter. The scale at which this maximum range occurs is indicated below the corresponding parameter name for each redshift. We note that, except for $n_\mathrm{s}$ and $\alpha_\mathrm{s}$, the smallest scales ($k > 5h/\text{Mpc}$), where uncertainties are larger, are excluded when determining $k_\mathrm{s}$.}
\end{figure*}

In Fig.~\ref{fig:param_sensitivity}, we show how the matter power spectrum varies with the cosmological parameters at $z=0$ and $z=2$. The reference cosmology is the same as that used in Ref.~\cite{Yang2025}. We see that large variations (up to $\sim 150\%$) in the matter power spectrum are supported by {\scriptsize GokuNEmu} around $\theta_\text{ref}$. While the results for $z=0$ are consistent with those in our previous work~\cite{Yang2025}, we notice that the current predictions are more accurate and smoother. For instance, {\scriptsize GokuEmu} failed to accurately generate matter power spectra for $w_a$ near the lower bound of its prior range (see panel 7 of Fig.~15 of Ref.~\cite{Yang2025}), while {\scriptsize GokuNEmu} successfully captures the dependence of the matter power spectrum on $w_a$ throughout the prior range.

From the figure, we also find that the most sensitive $k$ mode (denoted as $k_\text{s}$) at which the matter power spectrum varies with each of the parameters changes with redshift, typically shifting to higher $k$ as redshift increases. The values of $k_\text{s}$ are consistent with those reported in Ref.~\cite{Yang2025}, showing only minor differences that may arise from the different $k$-binning schemes and/or the wider $k$-range coverage adopted in this work. In addition, we notice that, for most of the parameters, the matter power spectrum is more sensitive to their changes at $z=2$ than at $0$ (thus wider fractional variations), except for $n_\text{s}$, $w_0$, and $w_a$.

\end{document}